\documentclass[twoside,english,sort&compress]{iopart}
\usepackage[T1]{fontenc}
\usepackage[utf8]{inputenc}
\usepackage{geometry}
\usepackage{color}
\usepackage{babel}
\usepackage{array}
\usepackage{amsbsy}
\usepackage{amstext}
\usepackage{graphicx}
\usepackage[numbers]{natbib}
\usepackage[unicode=true,
 bookmarks=true,bookmarksnumbered=false,bookmarksopen=false,
 breaklinks=false,pdfborder={0 0 1},backref=false,colorlinks=true]
 {hyperref}
\hypersetup{
 pdfauthor={Mr.wan},
  citecolor=blue,linkcolor=blue,urlcolor=blue}

\makeatletter

\providecommand{\tabularnewline}{\\}

\usepackage{iopams}
\usepackage{setstack}

\newcommand{\eqref}[1]{(\ref{#1})}
\usepackage[pagewise]{lineno}
\usepackage{tablefootnote}

\makeatother

\begin{document}
\title{EAST discharge prediction without integrating simulation results}
\author{Chenguang Wan$^{1,2,*}$, Zhi Yu$^{1}$, Alessandro Pau$^{3}$, Xiaojuan
Liu$^{1}$and Jiangang Li$^{1,2,*}$}
\address{1. Institute of Plasma Physics, Hefei Institutes of Physical Science,
Chinese Academy of Sciences, Hefei 230031, China}
\address{2. University of Science and Technology of China, Hefei, 230026, China}
\address{3. Eěcole Polytechnique Feědeěrale de Lausanne (EPFL), Swiss Plasma
Center (SPC), CH-1015 Lausanne, Switzerland}
\ead{\href{mailto:chenguang.wan@ipp.ac.cn}{chenguang.wan@ipp.ac.cn} and
\href{mailto:j_li@ipp.ac.cn}{j\_li@ipp.ac.cn}}
\begin{abstract}

In this work, a purely data-driven discharge prediction model was
developed and tested without integrating any data or results from
simulations. The model was developed based on the experimental data
from the Experimental Advanced Superconducting Tokamak (EAST) campaign
2010-2020 discharges and can predict the actual plasma current $I_{p}$,
normalized beta $\beta_{n}$, toroidal beta $\beta_{t}$, beta poloidal
$\beta_{p}$, electron density $n_{e}$, store energy $W_{mhd}$,
loop voltage $V_{loop}$, elongation at plasma boundary $\kappa$,
internal inductance $l_{i}$, q at magnetic axis $q_{0}$, and q at
95\% flux surface $q_{95}$. The average similarities of all the selected
key diagnostic signals between prediction results and the experimental
data are greater than 90\%, except for the $V_{loop}$ and $q_{0}$.
Before a tokamak experiment, the values of actuator signals are set
in the discharge proposal stage, with the model allowing to check
the consistency of expected diagnostic signals. The model can give
the estimated values of the diagnostic signals to check the reasonableness
of the tokamak experimental proposal.
\end{abstract}
\submitto{\NF }
\maketitle

\section{Introduction}

The entire prediction of a plasma discharge in a tokamak is a complicated
and critical task, which needs to be enhanced beyond the current capabilities
of available simulation codes. It is commonly used to check the consistency
of the modeled diagnostic signals, assist the experimental data analysis
phase, validate theoretical models, control technology R\&D, and provide
references for the design of an experiment. In the framework of conventional
discharge prediction, from a physics point of view, the primary method
is \textquotedbl Integrated Modeling\textquotedbl{} \citep{Falchetto2014}
derived from first principles. \textquotedbl Integrated Modeling\textquotedbl{}
involves lots of different physical processes in tokamaks. Its accuracy
depends on the completeness and consistency of the tokamak’s physics
derivations at the base of the model itself. High-fidelity and fast
simulation of the entire tokamak discharge is still an open problem
because of the high non-linearity, multi-spatial-temporal scales and
multi-physics nature of tokamak plasmas \citep{Bonoli2015}.

A neural network based method is an alternative approach for tokamak
discharge prediction without integrating complex physical modeling.
The method has been adopted in magnetic fusion research to solve a
variety of problems, including disruption prediction \citep{kates-harbeck2019be,Hu2021,Guo2021,Cannas2007,cannas2007support,Cannas2010,Yoshino2003},
electron temperature profile estimation from multi-energy SXR diagnostics
\citep{clayton2013electron}, radiated power estimation \citep{Barana2002},
filament detection \citep{CANNAS2019374}, simulation acceleration
\citep{Honda2019,Meneghini2017,Meneghini2021}, classifying confinement
regimes \citep{Murari2012}, plasma tomography \citep{Ferreira2020},
identification of instabilities \citep{MURARI20132}, estimation of
neutral beam effects \citep{Boyer2019}, determination of scaling
laws \citep{MURARI2010850,Gaudio_2014} coil current prediction with
the heat load pattern \citep{Bockenhoff2018}, equilibrium reconstruction
\citep{clayton2013electron,coccorese1994identification,Bishop1994,Jeon2001,Wang2016a,Joung2020},
and equilibrium solver \citep{Milligen1995}, control plasma \citep{Bishop1995,Yang_2020,Wakatsuki2019,Rasouli2013,yang2020modeling,Seo2021},
physic-informed machine learning \citep{Mathews2021}. Additionally,
a method mixed neural-network with simulation code for discharge prediction
\citep{wan2021} was noted. In that work \citep{wan2021}, only the
backward (past) observation information was used without considering
the forward (future) information. While there are tasks such as disruptions
prediction, which must inherently satisfy online causality settings,
other tasks like discharge predictive modeling can take advantage
of a wider context as far as offline analysis is concerned. In the
offline discharge prediction, where backward and forward information
is in principle equally important to describe the dynamic of the system.
Another limitation of that work is that it allows predicting only
a restricted set of signals (i.e., stored energy $W_{mhd}$, loop
voltage $V_{loop}$ and electron density $n_{e}$ ). The evolution
of a tokamak discharge is a complex process characterized by many
global parameters, such as equilibrium and kinetic quantities. The
restricted set of quantities namely $W_{mhd}$, $V_{loop}$, $n_{e}$,
is still not enough for the predictive modeling of a tokamak discharge.
Moreover, the model requires integrating a physical simulation code
to estimate the actual $I_{p}$ as an input signal. The total inference
time of that model would be long and inefficient. The Tokamak Simulation
Code \citep{Jardin_1993,Jardin1986} (TSC), as used in that paper
to simulate the entire discharge plasma current $I_{p}$ typically
takes several hours if multiple models (auxiliary-heating, current-drive,
alpha-heating, pellet-injection, etc) are included. In contrast, our
model's typical total inference time is \textasciitilde 1s. In the
present work, a new machine learning architecture was designed to
consider wider contextual information, predicted eleven key signals
of tokamak discharge, and values of all the input signals that can
be directly available or given by the machine learning model without
integrating any physical simulation code results.

In the present work, we trained the bidirectional long short-term
memory (LSTM) models using large-scale data from the EAST tokamak
\citep{Wan2015,Wan2013,Li2011} coming from 2010-2020 campaigns. Bidirectional
LSTM \citep{thireou2007,Schuster1997,Graves2005a} connects two hidden
layers with the information propagating in opposite directions to
the same output. The output layer can obtain information from backward
and forward states simultaneously with these designations. With the
96 actuator signals (introduce further in table \ref{tab:signals-list})
as input, the model is able to reproduce the whole tokamak discharge
time evolution of eleven key diagnostic signals, that are actual plasma
current $I_{p}$, normalized beta $\beta_{n}$, toroidal beta $\beta_{t}$,
beta poloidal $\beta_{p}$, electron density $n_{e}$, store energy
$W_{mhd}$, loop voltage $V_{loop}$, elongation at plasma boundary
$\kappa$, internal inductance $l_{i}$, q at magnetic axis $q_{0}$,
and q at 95\% flux surface $q_{95}$.

This paper is organized as follows. First, section \ref{sec:Model}
details the deep learning model architecture. Then, section \ref{sec:Dataset}
describes the data preprocessing and selection criteria. Section \ref{sec:Training}
shows the model training process. Next, section \ref{sec:Results}
presents the model results, and a depth analysis is given. Finally,
we make a brief discussion and conclusion in section \ref{sec:Conclusion}.

\section{Model \label{sec:Model}}

In this work, a bidirectional Recurrent Neural Network (RNN) \citep{thireou2007,Schuster1997,Graves2005a}
was developed, and contextual information was taken into account.
In this paper, our deep learning architecture stacked four bidirectional
LSTM cells.

Theoretically, the bidirectional LSTM network simultaneously minimizes
the objective function for the forward and the backward pass. Because
both future and past information is available to the model during
inference, the prediction of the model does not depend on an individually
defined delay parameter \citep{Schuster1997}. We consider the tokamak
discharge evolution as a sequence-to-sequence process, and as in a
language translation task, the bidirectional LSTM can utilize past
and future information for current prediction. In this work, the proposed
model is designed for the task of offline discharge modeling, and
it exploits a wider contextual information the future information
utilization is equivalent to relaxing the causality constraint to
obtain greater contextual information. It has been shown in several
works \citep{Abduljabbar2021,Siami-Namini2019} that a bidirectional
LSTM is often able to model more efficiently and robustly. Bidirectional
long-term dependencies between time steps of time series or sequence
data are particularly useful for regression tasks. The network, having
access to the complete time series at each time step, will be more
robust to the noise in the reconstruction of the tokamak discharge,
improving the similarity of the reconstructed parameters.

\begin{figure}
\begin{centering}
\includegraphics[width=0.7\textwidth]{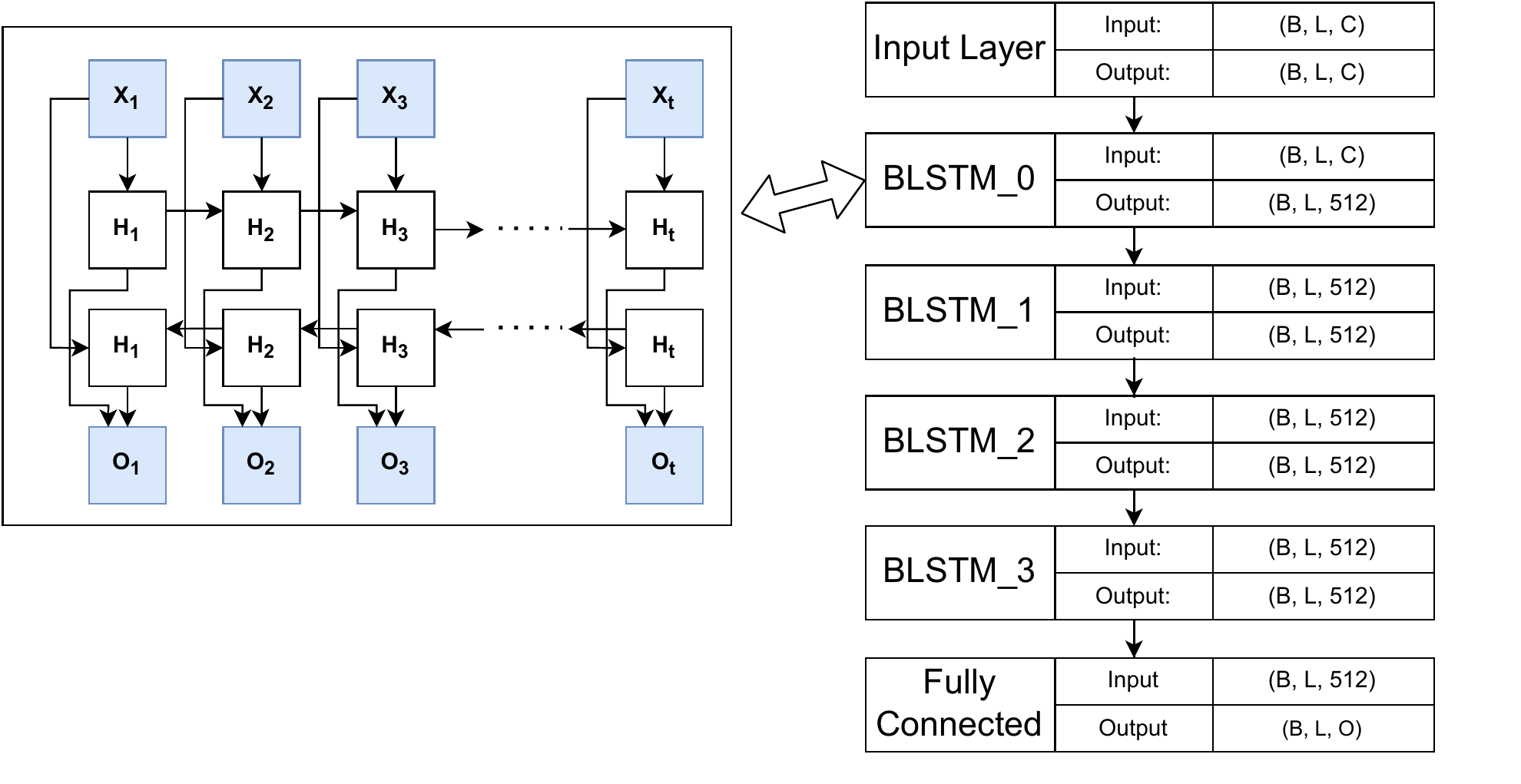}
\par\end{centering}
\caption{The architecture of bidirectional LSTM. The subfigure on the left
represents a bidirectional LSTM cell. Here, B, L, C, O are batch size,
the longest sequence length of a batch, feature size, and the number
of output signal channels. \label{fig:Architecture}}
\end{figure}

Figure \ref{fig:Architecture} shows the deep learning model architecture
stacked four bidirectional LSTMs. For any time step $t$, we define
the mini-batch input as $\boldsymbol{X_{t}}\in\mathbb{R}^{n\times d}$
(number of examples: $n$ , number of input features in each example:
$d$). In per layer bidirectional LSTM, the forward and backward hidden
states for this time step are $\overrightarrow{\boldsymbol{H}}_{t}\in\mathbb{R}^{n\times h}$
and $\overleftarrow{\boldsymbol{H}}_{t}\in\mathbb{R}^{n\times h}$
, respectively, where $h$ is the hidden units number. $\overrightarrow{\boldsymbol{H}}_{t}\in\mathbb{R}^{n\times h}$
and $\overleftarrow{\boldsymbol{H}}_{t}\in\mathbb{R}^{n\times h}$
are updated with standard Bidirectional LSTM layer operations \citep{Schuster1997}.

Next, the forward and backward hidden states $\overrightarrow{\boldsymbol{H}}_{t}$
and $\overleftarrow{\boldsymbol{H}}_{t}$ are concatenated to get
the hidden state $\boldsymbol{H}_{t}\in\mathbb{R}^{n\times2h}$ and
fed it into the next layer. When we consider dropout, the hidden state
$\boldsymbol{H}_{t}\in\mathbb{R}^{n\times2h}$ will be randomly masked
as zeros with a dropout rate $\delta$ of 0.1 at each step during
the \emph{training} phase. The dropout rate, as well as the other
hyperparameters, have been optimized maxing the performance of the
model on the validation set, according to the usual ``hyperparameter
tuning'' procedure. The dropout can easily help prevent overfitting
\citep{Srivastava2014}.

Last,output $\boldsymbol{O}_{t}\in\mathbb{R}^{n\times q}$(number
of outputs: q) is computed as follows:

\begin{equation}
\boldsymbol{O}_{t}=activation(\boldsymbol{H}_{t}\boldsymbol{W}_{hq}+\boldsymbol{b}_{q}).\label{eq:2}
\end{equation}

Here, the weight matrix $\boldsymbol{W}_{hq}\in\mathbb{R}^{2h\times q}$
and the bias $\boldsymbol{b}_{q}\in\mathbb{R}^{1\times q}$ are the
model parameters of the output layer.

One of the key features of a bidirectional LSTM is that information
from both directions of the sequence is used to estimate the current
time step. The architecture uses future (forward) and past (backward)
information to perform the inference. The final layer activation function
is a linear function since we are dealing with a regression task and
don't want to constrain output parameters values.

\section{Dataset \label{sec:Dataset}}

The whole EAST's data system stores more than 3000 raw channel data
and thousands of processed physical analysis data \citep{wang2018studyof},
which records the whole process of the EAST discharge. All the tokamak
data was divided into three categories: configuration parameters,
actuator signals, and diagnostic signals. The configuration parameters
describe constants related to the tokamak such as magnetic probe positions.
The actuator signals such as the power of ECRH, etc., are actively
controlled quantities. The diagnostic signals are observable parameters
passively measured from internal plasma such as loop voltage $V_{loop}$
etc. The configuration parameters are fixed during the tokamak experiments.
The discharge prediction task can be essentially reduced to mapping
actuator (input) signals to diagnostic (output) signals.

In the present work, the output signals include all important 0-D
EAST diagnostic signals routinely available after a discharge. The
input signals include all the quantities that may affect the output.
Table \ref{tab:signals-list} contains detailed information about
input and output signals. The input signal ``Ref. Shape'', in a
particular, requires a more in depth discussion. According to the
tokamak magnetic control system workflow \citep{DeTommasi2019}, the
shape references affect the in-vessel Rogowski coil current (IC1).
So when estimating IC1, such shape references data is required. On
EAST, shape feedback control is realized using the isoflux control
scheme \citep{Anand2021b}. As the key concept of isoflux control,
the plasma target shape (Ref. Shape) is interpreted as a set of control
points defining the desired plasma boundary, and the flux at each
control point is regulated to be equal. Usually, one of the control
points is chosen as a reference point, which is typically a point
on the limiter in a limited plasma configuration or the X point in
a diverted configuration. Thus the plasma boundary is controlled by
adjusting the PF coil currents to eliminate the flux error between
the reference point and other boundary control points. Some of the
selected signals are processed signals with clear physical meaning,
and others are unprocessed raw electrical signals. Since some signals
of the actual EAST experimental diagnostic system are not processed,
we directly selected some unprocessed raw electrical signals. As long
as the input signals include information to determine the output,
the unprocessed signals will not affect the modeling result.
\begin{center}
\begin{table}
\caption{The list of signals. The \textquotedblleft raw signal\textquotedblright{}
means the original electrical signal, and these could be converted
to signals with physical meaning. The IC1 was estimated using a machine
learning model and then fed to the diagnostic signals (output) reconstruction
model as input.\label{tab:signals-list}}

\begin{tabular}[t]{>{\raggedright}p{0.05\paperwidth}>{\raggedright}p{0.2\paperwidth}>{\raggedright}p{0.07\paperwidth}>{\raggedright}p{0.07\paperwidth}>{\raggedright}p{0.05\paperwidth}>{\raggedright}p{0.25\paperwidth}}
\hline 
\raggedright{}{\footnotesize{}Signals} & \raggedright{}{\footnotesize{}Physics meanings} & \raggedright{}{\footnotesize{}Unit} & \raggedright{}{\footnotesize{}Number of Channels} & \raggedright{}{\footnotesize{}Sampling rate} & \raggedright{}{\footnotesize{}Meaning of Channels}\tabularnewline
\hline 
\multicolumn{3}{l}{{\footnotesize{}Output Signals}} & \raggedright{}{\footnotesize{}12} & \raggedright{} & \raggedright{}\tabularnewline
\hline 
\raggedright{}{\footnotesize{}$\text{Act.}I_{p}$} & \raggedright{}{\footnotesize{}Actual plasma current} & \raggedright{}{\footnotesize{}$A$} & \raggedright{}{\footnotesize{}1} & \raggedright{}{\footnotesize{}$1kHz$} & \raggedright{}{\footnotesize{}Actual plasma current}\tabularnewline
\raggedright{}{\footnotesize{}$n_{e}$} & \raggedright{}{\footnotesize{}Electron density} & \raggedright{}{\footnotesize{}$10^{19}m^{-3}$} & \raggedright{}{\footnotesize{}1} & \raggedright{}{\footnotesize{}$1kHz$} & \raggedright{}{\footnotesize{}Electron density}\tabularnewline
\raggedright{}{\footnotesize{}$\ensuremath{W_{mhd}}$} & \raggedright{}{\footnotesize{}Plasma stored energy} & \raggedright{}{\footnotesize{}$J$} & \raggedright{}{\footnotesize{}1} & \raggedright{}{\footnotesize{}$20Hz$} & \raggedright{}{\footnotesize{}Plasma stored energy}\tabularnewline
\raggedright{}{\footnotesize{}$\text{\ensuremath{V_{loop}}}$} & \raggedright{}{\footnotesize{}Loop voltage} & \raggedright{}{\footnotesize{}$V$} & \raggedright{}{\footnotesize{}1} & \raggedright{}{\footnotesize{}$1kHz$} & \raggedright{}{\footnotesize{}Loop voltage}\tabularnewline
\raggedright{}{\footnotesize{}$\ensuremath{\beta_{n}}$} & \raggedright{}{\footnotesize{}Normalized beta} & \raggedright{}{\footnotesize{}dimensionless} & \raggedright{}{\footnotesize{}1} & \raggedright{}{\footnotesize{}$15Hz$} & \raggedright{}{\footnotesize{}Normalized beta}\tabularnewline
\raggedright{}{\footnotesize{}$\ensuremath{\beta_{t}}$} & \raggedright{}{\footnotesize{}Toroidal beta} & \raggedright{}{\footnotesize{}dimensionless} & \raggedright{}{\footnotesize{}1} & \raggedright{}{\footnotesize{}$15Hz$} & \raggedright{}{\footnotesize{}Toroidal beta}\tabularnewline
\raggedright{}{\footnotesize{}$\ensuremath{\beta_{p}}$} & \raggedright{}{\footnotesize{}Beta poloidal} & \raggedright{}{\footnotesize{}dimensionless} & \raggedright{}{\footnotesize{}1} & \raggedright{}{\footnotesize{}$15Hz$} & \raggedright{}{\footnotesize{}Beta poloidal}\tabularnewline
\raggedright{}{\footnotesize{}$\kappa$} & \raggedright{}{\footnotesize{}Elongation at plasma boundary} & \raggedright{}{\footnotesize{}dimensionless} & \raggedright{}{\footnotesize{}1} & \raggedright{}{\footnotesize{}$15Hz$} & \raggedright{}{\footnotesize{}Elongation at plasma boundary}\tabularnewline
\raggedright{}{\footnotesize{}$l_{i}$} & \raggedright{}{\footnotesize{}Internal inductance} & \raggedright{}{\footnotesize{}dimensionless} & \raggedright{}{\footnotesize{}1} & \raggedright{}{\footnotesize{}$15Hz$} & \raggedright{}{\footnotesize{}Internal inductance}\tabularnewline
\raggedright{}{\footnotesize{}$q_{0}$} & \raggedright{}{\footnotesize{}q at magnetic axis} & \raggedright{}{\footnotesize{}dimensionless} & \raggedright{}{\footnotesize{}1} & \raggedright{}{\footnotesize{}$15Hz$} & \raggedright{}{\footnotesize{}q at magnetic axis}\tabularnewline
\raggedright{}{\footnotesize{}$q_{95}$} & \raggedright{}{\footnotesize{}q at 95\% flux surface} & \raggedright{}{\footnotesize{}dimensionless} & \raggedright{}{\footnotesize{}1} & \raggedright{}{\footnotesize{}$15Hz$} & \raggedright{}{\footnotesize{}q at 95\% flux surface}\tabularnewline
\hline 
\multicolumn{3}{l}{{\footnotesize{}Feedback Signal}} & \raggedright{}{\footnotesize{}1} & \raggedright{} & \raggedright{}\tabularnewline
\hline 
\raggedright{}{\footnotesize{}IC1} & \raggedright{}{\footnotesize{}In-vessel coil no.1 current} & \raggedright{}{\footnotesize{}$A$} & \raggedright{}{\footnotesize{}1} & \raggedright{}{\footnotesize{}$1kHz$} & \raggedright{}{\footnotesize{}In-vessel Rogowski coil no.1 current}\tabularnewline
\hline 
\multicolumn{3}{l}{{\footnotesize{}Input Signals}} & \multicolumn{3}{l}{{\footnotesize{}95}}\tabularnewline
\hline 
\raggedright{}{\footnotesize{}Ref.$I_{p}$} & \raggedright{}{\footnotesize{}Reference plasma current} & \raggedright{}{\footnotesize{}$A$} & \raggedright{}{\footnotesize{}1} & \raggedright{}{\footnotesize{}$1kHz$} & \raggedright{}{\footnotesize{}Reference plasma current}\tabularnewline
\raggedright{}{\footnotesize{}PF} & \raggedright{}{\footnotesize{}Current of Poloidal field (PF) coils} & \raggedright{}{\footnotesize{}$A$} & \raggedright{}{\footnotesize{}14} & \raggedright{}{\footnotesize{}$1kHz$} & \raggedright{}{\footnotesize{}PF 0-14 current}\tabularnewline
\raggedright{}{\footnotesize{}$B_{t0}$} & \raggedright{}{\footnotesize{}Toroidal magnetic field} & \raggedright{}{\footnotesize{}$T$} & \raggedright{}{\footnotesize{}1} & \raggedright{}{\footnotesize{}$1kHz$} & \raggedright{}{\footnotesize{}Toroidal field at magnetic axis}\tabularnewline
\raggedright{}{\footnotesize{}LHW} & \raggedright{}{\footnotesize{}Power of Lower Hybrid Wave Current Drive
and Heating System} & \raggedright{}{\footnotesize{}$kW$} & \raggedright{}{\footnotesize{}4} & \raggedright{}{\footnotesize{}$20kHz$} & \raggedright{}{\footnotesize{}2.45 GHz LHW, and 4.6 GHz LHW}\tabularnewline
\raggedright{}{\footnotesize{}NBI} & \raggedright{}{\footnotesize{}Neutral Beam Injection System} & \raggedright{}{\footnotesize{}Raw signal} & \raggedright{}{\footnotesize{}8} & \raggedright{}{\footnotesize{}$5kHz$} & \raggedright{}{\footnotesize{}Acceleration voltage and beam current,
of No. 1-2 left/right ion source.}\tabularnewline
\raggedright{}{\footnotesize{}ICRH} & \raggedright{}{\footnotesize{}Ion Cyclotron Resonance Heating System} & \raggedright{}{\footnotesize{}Raw signal} & \raggedright{}{\footnotesize{}16} & \raggedright{}{\footnotesize{}$5kHz$} & \raggedright{}{\footnotesize{}Output of detector for rejected power
of No. 1-16 transmitter}\tabularnewline
\raggedright{}{\footnotesize{}ECRH/ ECCD} & \raggedright{}{\footnotesize{}Electron Cyclotron Resonance Heating/Current
Drive System} & \raggedright{}{\footnotesize{}Raw signal} & \raggedright{}{\footnotesize{}4} & \raggedright{}{\footnotesize{}$50kHz$} & \raggedright{}{\footnotesize{}Output of detector for injected power
measurement No. 1-4 gyrotron}\tabularnewline
\raggedright{}{\footnotesize{}GPS} & \raggedright{}{\footnotesize{}Gas Puffing System} & \raggedright{}{\footnotesize{}Raw signal} & \raggedright{}{\footnotesize{}12} & \raggedright{}{\footnotesize{}$10kHz$} & \raggedright{}{\footnotesize{}Horizontal ports J, K, D, B; Upper port
O; Lower ports O, C, H}\tabularnewline
\raggedright{}{\footnotesize{}SMBI} & \raggedright{}{\footnotesize{}Supersonic Molecular Beam Injection} & \raggedright{}{\footnotesize{}Raw signal} & \raggedright{}{\footnotesize{}3} & \raggedright{}{\footnotesize{}$10kHz$} & \raggedright{}{\footnotesize{}3 ports of SMBI}\tabularnewline
\raggedright{}{\footnotesize{}PIS} & \raggedright{}{\footnotesize{}Pellet Injection System} & \raggedright{}{\footnotesize{}Raw signal} & \raggedright{}{\footnotesize{}1} & \raggedright{}{\footnotesize{}$10kHz$} & \raggedright{}{\footnotesize{}1 injection line for Pellet Injection}\tabularnewline
{\footnotesize{}Ref. Shape} & {\footnotesize{}Shape reference} & {\footnotesize{}Raw signal} & {\footnotesize{}31} & {\footnotesize{}$1kHz$} & {\footnotesize{}20 groups of control points}\tabularnewline
\hline 
\end{tabular}
\end{table}
\par\end{center}

Tokamak discharge evolution is a multi-spatial-temporal scales, non-linear,
multi-physics quantities coupling process. There is no simple function
to determine the relationship between the actuator signals and the
diagnostic signals. Different diagnostic parameters are determined
by different inputs. In the present work, the input signals can not
simply and directly determine all the output signals. In our model
training experiments, $l_{i}$, $\beta_{p}$ can not be accurately
estimated by the input signals that do not include IC1. When the input
signals include the actual IC1, all output signals of the present
work can be estimated. Since IC1 values cannot be obtained through
direct measurement, they have been indirectly estimated through a
machine learning model.

\section{Training \label{sec:Training}}

This section shows details of model training and data processing.
The training model is divided into four steps as follows:
\begin{enumerate}
\item Obtaining and resampling the data of the 108 data channels (including
input, feedback, and output signals as shown in table \ref{tab:signals-list})
from the EAST source database, then storing it in HDF5 \footnote{Hierarchical Data Format (HDF) is a set of file formats (HDF4, HDF5)
designed to store and organize large amounts of data.} file with each HDF5 file data properties (standard deviation $\delta$,
mean $\mu$, etc.) stored in MongoDB \citep{mongodb}.
\item Standardizing the data with z-scores \footnote{z-score is calculated by $z=(x-\mu\text{)/\ensuremath{\sigma}}$ where
$\mu$ is the mean of the population. $\sigma$ is the standard deviation
of the population.} (also known as standard scores).
\item Using bucketing (explained further in section \ref{subsec:Bucketing})
batch training for the deep learning model to be reconstructed IC1.
\item Integrating the estimated IC1 as input for the eleven key diagnostic
signals (table \ref{tab:signals-list} output signals) reconstruction
model training.
\end{enumerate}

\subsection{Obtaining and resampling}

The dataset is selected from EAST tokamak during the 2010-2020 campaigns,
and the discharge shot number is in the range \#14866-88283 \citep{Wan2015,Wan2013,Li2011}.
A total of 26230 normal shots were selected. ``Normal shot'' means
that the plasma current is safely landed without disrupting and the
flat-top duration is of at least two seconds. Moreover all the key
quantities (i.e., magnetic field $B_{t}$, PF) describing the magnetic
configuration as well as the actual plasma current are available for
the entire duration of the shots. If there is not a certain magnetic
field configuration, it is impossible to constrain (or control) the
plasma. Furthermore, during a tokamak experiment, the actual $I_{p}$
is a key physical quantity, the experiment is generally considered
a failure if there is no actual $I_{p}$. In model usage, actual $I_{p}$
is a output signal. Additionally, from a pure technical point of view,
the training as well as the testing of the deep learning model cannot
be performed if unless at least one output signal is available. Three
different data sets are needed while developing the bidirectional
LSTM model for discharge modeling. A training set is required for
training the model. A validation set is needed for hyper-parameters
tuning, whereas a test set is used to measure the final accuracy of
the deep learning model, as well as its generalization capability.
The training set should use the earliest data of the selected EAST
campaign to simulate the new data faced in practice, as it always
happens in practice. It is worth mentioning that, despite of the chronologically
order, validation and test sets do not contain more recent pulse scheduling/piloting
techniques with respect to what is not included in the training set.
The capability of more advanced operations and control schemes as
well as the target plasma evolve over the years according to the experimental
program. For the sake of consistency, completely new control techniques
and piloting schemes have not been considered for the evaluation of
the current models. The test set and validation set are assumed to
be statistically similar, so we can expect the best performance on
the test set by optimizing the accuracy of the validation set. For
each epoch, the model must input every shot in the validation set
to get the performance metric, and the validation set is not used
to update the model parameters. So relatively small validation set
can accelerate the model training. The shot range of the validation
set and the test set should not have any intersection to ensure the
objectivity of the test set. The test set should only be used once,
that is for assessing the real accuracy of the model. In other words,
the test set should be used only for model testing and should not
be present during the model training or tuning phase. These three
data sets have been carefully selected to meet these strict requirements.
Shot \#14866-74999 in the EAST database is selected as the training
set. 21192 normal shots are reserved at last. The validation set has
shot numbers in the range \#75000-77000. Shots \#77000-88283 are in
the test set.

All data are resampled with the same sample rate $1kHz$. Although
we used a relatively low sampling rate, the original resampled data
set still contains 55GB of data. Therefore for each shot, the data
is saved to an HDF5 file, not in the database server for quick and
robust training. The metadata is stored in MongoDB for double-checking
the data validity and availability by the human and program.

We align our data with the same time axis by linearly interpolating
to up-sample and a simple moving average (SMA) to down-sample. The
simple moving average (SMA) is the unweighted mean of the \textbf{$k$}
data-points and works like a low-pass filter. Our SMA used the information
from the later time. Since we use a bidirectional LSTM as architecture
for the model and we aim to offline discharge modeling, we can remove
the causality constraint for the sequences fed to the model for training.
From another perspective, since high-frequency fluctuations are not
a relevant outcome of the experimental proposal stage, we process
the data by filtering out high-frequency fluctuations.

\subsection{Data standardization}

Firstly, all source data was saved in the HDF5 file. Then, discharge
duration time and every signal mean, variance, and existence flag
will be saved to the MongoDB database shot-by-shot. Saving the mean
and variance of each signal of each shot data is necessary not only
to double-check the presence of outliers in the signals but also for
calculating the global means and standard deviations for the huge
dataset by MapReduce \citep{Dean2008}. If a signal in a shot has
outliers, then the signal values in this shot will not be used to
calculate the signal's global mean and standard deviation. MapReduce
is a programming model and an associated implementation for processing
and generating big data sets with a parallel, distributed algorithm
on a cluster. The reasons for using MapReduce are that if the global
mean and standard deviation were calculated directly, the calculation
would be overflowed. When the source data means and standard deviations
have been calculated, the z-scores will be applied for standardization.
In this step, if the input data have NaN (invalid value) or Inf (infinite
value) will be replaced by a linear interpolation value and $3.2\text{\ensuremath{\times10^{32}}}$(it
is not the maximum value of the float 32 type, but it is large enough.
And it can still be calculated without overflowing), respectively.

\subsection{Bucketing batchwise training. \label{subsec:Bucketing}}

Mini-batching gradient descent \citep{dean2012large,Huang2013} is
a valuable technique for enhancing GPU performance \citep{Chetlur2014}
and accelerating the training convergence of deep-learning models.
The parameter loss gradients are computed for several examples in
parallel and then averaged. However, this operation requires all examples
have the same length. Therefore, training RNN or its variants on a
large amount of data with different lengths is a quite challenging
problem. Bucketing \citep{Khomenko2016} was used to solve this problem
in the present work.

The bucketing method can be reduced to a partition problem. Let $S=\left\{ s_{1},s_{2},\ldots,s_{n}\right\} $
be sequences set and the length of sequence $s_{i}$ is $l_{i}=\left|s_{i}\right|$.
Each GPU processes sequences in a mini-batch in a synchronized parallel
manner, so processing time of a mini-batch $I_{\text{batch}}=\left\{ s_{1},s_{2},\ldots,s_{k}\right\} $
is proportional to $O\left(\text{max}_{i\in1,\ldots,k}l_{i}\right)$
and processing time of the whole set is expressed as:

\begin{equation}
T(S)=O\left(n/k*\text{max}_{i\in1,\ldots,k}l_{i}\right).\label{eq:1-1}
\end{equation}

\begin{figure}
\begin{centering}
\includegraphics[width=0.5\columnwidth]{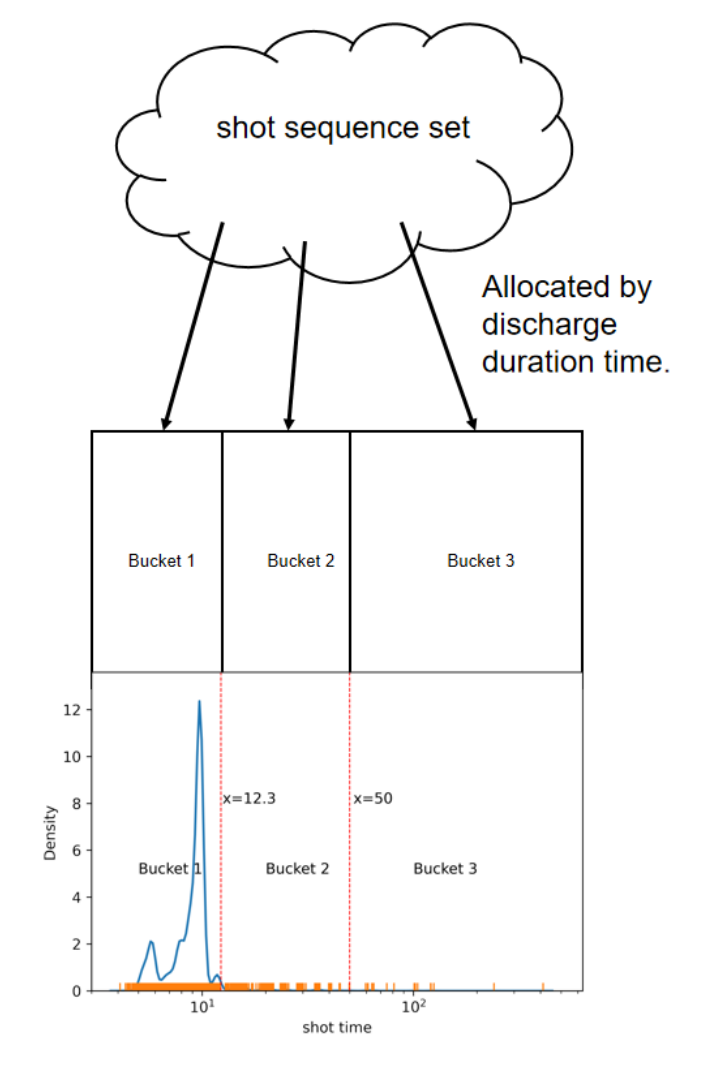}
\par\end{centering}
\caption{The shots during time distribution of the input discharge sequences.
The number of buckets $B=3$. The y-axis represents distribution density,
normalized as a probability density function. \label{fig:Bucketing}}
\end{figure}

If the dataset sequences were shuffled randomly before splitting,
the minimum and maximum sequence lengths in the mini-batch would be
very different. As a result, the GPU would do additional work for
processing the meaningless tails of shorter sequences. Additionally,
the too-long sequence with a non-suitable batch size will be overflowed
due to the GPU memory capacity limit. Specifically, if we use the
same batch size for long sequences input as for short sequences input,
it will take up more GPU memory which will easily cause overflows.
We used customized bucketing to optimize the batch training to overcome
this flaw and reduce training time. As shown in figure \ref{fig:Bucketing},
according to length, all sequences are partitioned to $B$ buckets.
Let $S_{i}=\left\{ s_{j_{1}},s_{j_{2}},\ldots s_{j_{k_{i}}}\right\} $.
For each bucket, we execute the mini-batch training with different
batch sizes. The processing time of the whole set is expressed as:

\begin{equation}
T(S)=\sum_{i=1}^{B}O(T(S_{i})).\label{eq:}
\end{equation}

We manually partitioned the sequence length set in the present work
because the different sequence length sets will use different batch
sizes. The result of the partition is shown in figure \ref{fig:Bucketing}.
Bucket 1,2,3 are in intervals $\left[2,12.3\right]$, $\left(12.3,50\right]$,
and $\left(50,412\right]$ respectively. Because of the GPU memory
capacity limit, the batch size of the three buckets is set at 8,4,1,
respectively. These batch sizes can control GPU memory overflow and
easily allocate the input tensor for each GPU.

The sequences within every bucket were shuffled randomly. And then,
the sequences were generated batch-by-batch. To train batchwise with
a batch size $M$, we need M independent shot discharge sequences
of the same bucket to feed to the GPU. The different length discharge
sequences were padded by zeros to the same length. We did this by
using M processes to read sequence data in parallel. The M sequences
were fed to a buffer first to solve the problem of GPU and CPU speed
mismatch since data from HDF5 files are read through a CPU.

\subsection{Model training \label{subsec:Model-training}}

IC1 is an in-vessel Rogowski coil current \citep{Xiao2008,DeTommasi2019},
and this current does not have a direct reference signal. In-vessel
current distribution play a key role in the accuracy of reconstruction
through a free boundary plasma equilibrium solver. The estimation
of such in-vessel current distribution through simulation codes is
very expensive and given the large number of discharges used to train
and assess the performance of the model described in this paper, we
decided to rely on the IC1 measured current, which acts basically
as a proxy for the in-vessel current distribution. Although this might
potentially introduce errors in the reconstruction of equilibrium
quantities, it was found that not including IC1 in the set of input
parameters heavily affected the accuracy of the reconstruction of
quantities such as $\beta_{n}$, $\beta_{p}$ and $l_{i}.$ The IC1
signal can not be programmed or manually set in the experiment proposal
stage. Therefore, first, we train a machine learning model (same model
architect as depicted in figure \ref{fig:Architecture}) to reconstruct
IC1. Secondly, we connected the trained IC1 reconstruction model and
the diagnostic parameters reconstruction model for training, where
the IC1 reconstruction part was fine-tuned using a minimal learning
rate. For this reason, the output of the IC1 reconstruction model
and the input IC1 to the deep learning model are different.

To facilitate fine-tuning the IC1 part, the model for reconstructing
IC1 and the diagnostic parameters has the same architecture, initializer,
optimizer, training set, validation set, test set, etc, in which only
the input and output are different. The input of the IC1 reconstruction
includes all input signals in table \ref{tab:signals-list}. The input
of the diagnostic parameters reconstruction includes the output of
the IC1 reconstruction model and does not use the shape reference
signals. The training process is similar to using the trained model,
as shown in figure \ref{fig:Using}.

The deep learning model uses end-to-end training executed on 8x Nvidia
P100 GPUs with PyTorch \citep{NEURIPS2019_9015} on the Centos7 operating
system. The weight initialization scheme for the deep learning model
is Xavier initialization \citep{Glorot2010}, bias initializer is
zeros, and optimizer is RMSprop \citep{Graves2013}. The loss function
of this training should be noted. The function is Masked MSELoss,
which has some improvements for mean squared error (MSE) loss function.
The MaskedMSEloss can be described as:

\begin{equation}
l\left(\boldsymbol{x},\boldsymbol{y}\right)=L=\frac{\sum_{i=0}^{i=N}\{l_{1},l_{2},\ldots,l_{N}\}}{N},\label{eq:-1}
\end{equation}

\begin{equation}
l_{i}=\sum_{j=0}^{j=\text{len}}\left(\boldsymbol{x}_{j}^{i}-\boldsymbol{y}_{j}^{i}\right)^{2},\label{eq:-2}
\end{equation}

where N is the batch size, $\boldsymbol{x}$ and $\boldsymbol{y}$
are the batch experimental sequence and batch predicted sequence,
$\boldsymbol{x}_{j}^{i}$, $\boldsymbol{y}_{j}^{i}$ are the $j$th
point values of the $i$th experimental sequence and predicted sequence.
The subtlety of this work is “$j=\text{len}$”, where ``len'' is
the length of the $i$th sequence. So the masked MSE function can
prevent useless training of the zeros padding section of the sequence.
In training our model, multiple sets of hyperparameters were tried.
It was a trial-and-error approach on different sets of hyperparameters.
The best hyperparameter set was selected based on the best performances
of the validation set. Table \ref{tab:Hype} shows the best hyperparameter
set.

\begin{table}
\begin{centering}
\caption{Our model Hyperparameters \label{tab:Hype}}
\par\end{centering}
\centering{}%
\begin{tabular}{lll}
\hline 
Hyperparameter & Explanation & Best value\tabularnewline
\hline 
$\eta$ & Normal learning rate & $3\times10^{-3}$ or $5\times10^{-3}$\tabularnewline
$\eta_{f}$ & Fine-tuning learning rate & $1\times10^{-4}$\tabularnewline
$\gamma$ & Momentum factor & 0.5\tabularnewline
L2 & L2 regularization rate & 0.01\tabularnewline
Loss function & Loss function type & Masked MSE\tabularnewline
Optimizer & Optimization scheme & RMSprop\tabularnewline
Dropout & Dropout probability & 0.1\tabularnewline
dt & Time step & 1ms\tabularnewline
Batch\_size & Batch size & 8,4,1\tabularnewline
RNN\_type & Type of RNN & Bidirectional LSTM\tabularnewline
$n_{rnn}$ & Number of RNN stacked & 4\tabularnewline
$H_{rnn}$ & Hidden size of RNN & 512\tabularnewline
$n_{encoder}$ & Number of BiLSTMs stacked in encoder & 2\tabularnewline
$n_{decoder}$ & Number of BiLSTMs stacked in decoder & 2\tabularnewline
\hline 
\end{tabular}
\end{table}

\section{Results \label{sec:Results}}

\begin{figure}
\begin{centering}
\includegraphics[width=0.7\columnwidth]{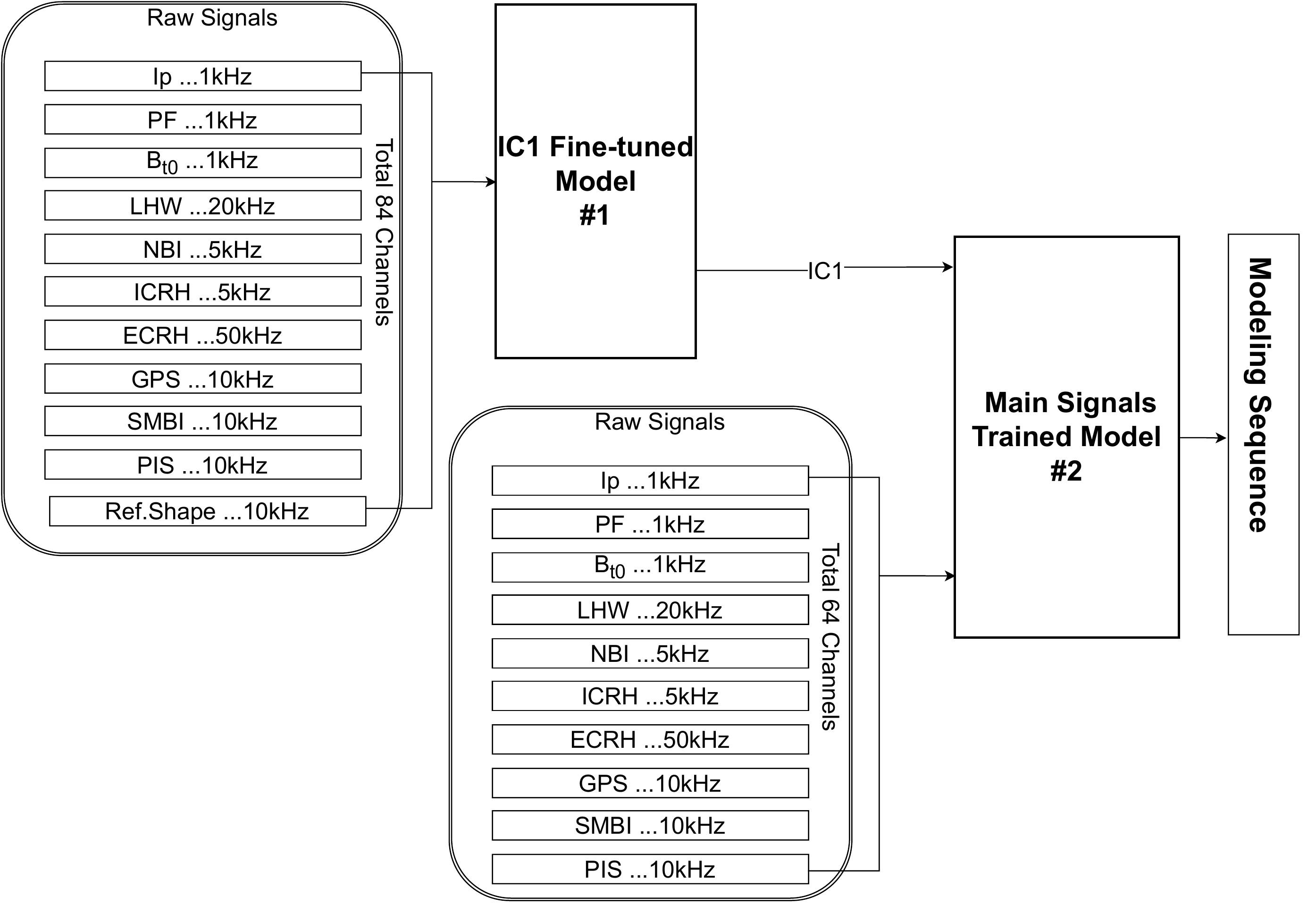}
\par\end{centering}
\caption{Using the trained model. \label{fig:Using}}
\end{figure}

The model was tested on unseen data (test set), as shown in figure
\ref{fig:Using}. In the present work, models are used in sequence.
The \#1 deep learning model is used first to get the IC1 estimated
values and then the \#2 deep learning model to estimate the tokamak's
main diagnostic signals. The preprocessing steps on new data are the
same as before, keeping the training set parameters (the mean $\mu$
and the standard deviation $\sigma$) for standardization.

This section details the results and analyzes. The results include
representative modeling results and eleven diagnostic signal similarity
distributions of the test set. Additionally, the present work uses
similarity and means square error (MSE) as quantitative measurements
of the accuracy of the modeling results.

\begin{equation}
S\left(\boldsymbol{x},\boldsymbol{y}\right)=\max\left(\frac{\Sigma(\boldsymbol{x}-\bar{\boldsymbol{x}})(\boldsymbol{y}-\bar{\boldsymbol{y}})}{\sqrt{\Sigma(\boldsymbol{x}-\bar{\boldsymbol{x}})^{2}\Sigma(\boldsymbol{y}-\bar{\boldsymbol{y}})^{2}}},0\right),\label{eq:3}
\end{equation}

\begin{equation}
MSE\left(\boldsymbol{x},\boldsymbol{y}\right)=\frac{1}{n}\sum_{i=1}^{n}(\boldsymbol{x}_{i}-\boldsymbol{y}_{i})^{2},\label{eq:4}
\end{equation}

where $\boldsymbol{x}$ is experimental data, $\boldsymbol{y}$ is
modeling result, $\bar{\boldsymbol{x}}$, $\bar{\boldsymbol{y}}$
are the means of the vector $\boldsymbol{x}$ and vector $\boldsymbol{y}$,
$\boldsymbol{x}_{i}$, $\boldsymbol{y}_{i}$ are the point values
of the vector $\boldsymbol{x}$, $\boldsymbol{y}$. The MSE is the
mean ($\frac{1}{n}\sum_{i=1}^{n}$) of the squares of the errors $(\boldsymbol{x}_{i}-\boldsymbol{y}_{i})^{2}$.
MSE will be affected by the outlier but it can more accurately measure
the difference of values. Similarity can only measure whether the
trend is consistent, but it cannot measure the difference in value.

\begin{figure}
\begin{centering}
\includegraphics[height=0.85\textheight]{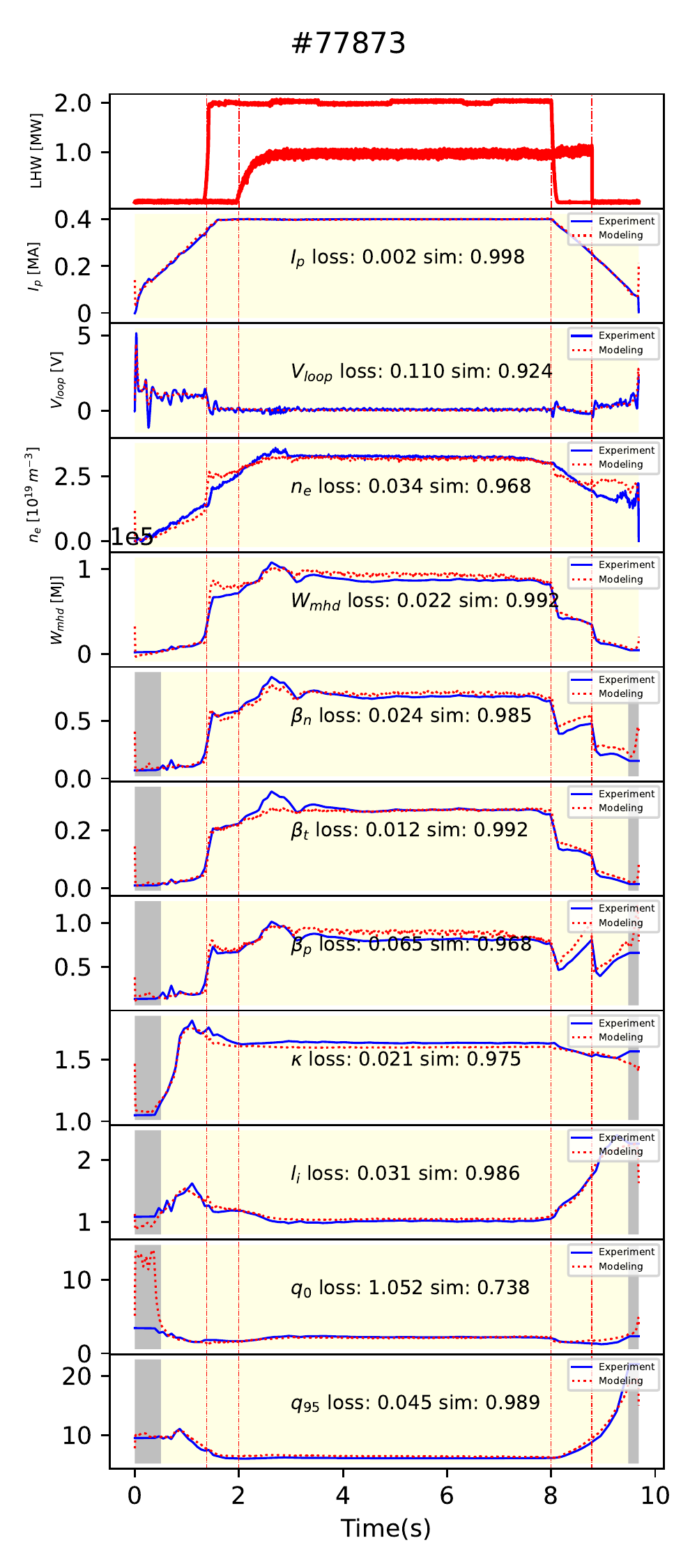}
\par\end{centering}
\caption{Comparison of the modeling result and EAST experiment data. \textquotedblleft sim\textquotedblright{}
is similarity (equation \ref{eq:3}) and \textquotedblleft loss\textquotedblright{}
is MSE (equation \ref{eq:4}). Shot \#77873 have two LHW injections
during discharge. No experimental data are available for the gray
area, so the modeling results for this area are unreliable. During
model training, we used the start value and end value to fill these
areas for time axis alignment. \label{fig:Demo_out}}
\end{figure}

A typical entire process discharge prediction of our model is shown
in figure \ref{fig:Demo_out}. In the present work, the trained model
can reproduce the eleven diagnostic signals during experimental proposal
stage, from ramp-up to ramp-down, without relying on any physical
codes.

Figure \ref{fig:Demo_out} shows our model can accurately reproduces
the slopes of the ramp-up and ramp-down and the amplitude of the flat-top.
The model can also reflect the external auxiliary system signal impact
on the diagnostic signal. The vertical dash-dot lines of figure \ref{fig:Demo_out}
indicate the rising and falling edges of the external auxiliary system
signals and the plasma response.

\begin{figure}
\begin{centering}
\includegraphics[width=0.85\columnwidth]{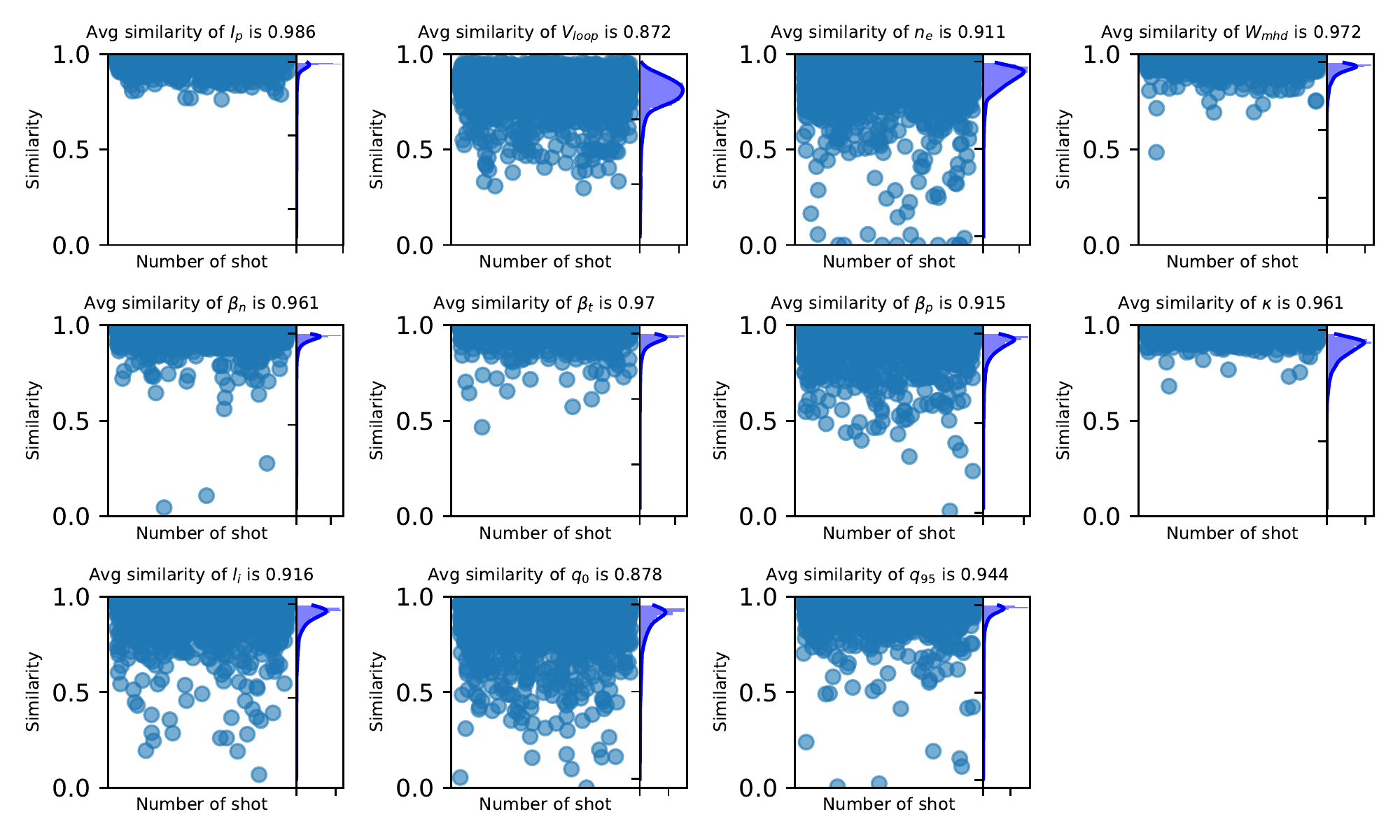}
\par\end{centering}
\caption{The similarity distribution and average similarity in the whole test
set. The figure shows the similarity distributions of the output signals
(see table \ref{tab:signals-list}) \label{fig:distribution}}
\end{figure}

\begin{table}
\begin{centering}
\caption{The average similarity and MSE of the eleven diagnostic signals. \label{tab:average}}
\par\end{centering}
\centering{}%
\begin{tabular}{>{\raggedright}p{0.1\paperwidth}ll}
\hline 
\raggedright{}Output signals & Average similarity & Average MSE\tabularnewline
\hline 
\raggedright{}{\footnotesize{}$\text{Act.}I_{p}$} & 0.986 & 0.0015\tabularnewline
\raggedright{}{\footnotesize{}$n_{e}$} & 0.911 & 0.341\tabularnewline
\raggedright{}{\footnotesize{}$\ensuremath{W_{mhd}}$} & 0.972 & 0.077\tabularnewline
\raggedright{}{\footnotesize{}$\text{\ensuremath{V_{loop}}}$} & 0.872 & 0.354\tabularnewline
\raggedright{}{\footnotesize{}$\ensuremath{\beta_{n}}$} & 0.961 & 0.062\tabularnewline
\raggedright{}{\footnotesize{}$\ensuremath{\beta_{t}}$} & 0.970 & 0.130\tabularnewline
\raggedright{}{\footnotesize{}$\ensuremath{\beta_{p}}$} & 0.915 & 0.179\tabularnewline
\raggedright{}{\footnotesize{}$\kappa$} & 0.961 & 0.0019\tabularnewline
\raggedright{}{\footnotesize{}$l_{i}$} & 0.916 & 0.127\tabularnewline
\raggedright{}{\footnotesize{}$q_{0}$} & 0.878 & 0.129\tabularnewline
\raggedright{}{\footnotesize{}$q_{95}$} & 0.944 & 0.093\tabularnewline
\hline 
\end{tabular}
\end{table}

The whole test data set of shot range 77000-88283 is used to quantitatively
evaluate the reliability of the modeling results of the eleven signals.
The statistical results of the similarity and MSE between modeling
results and experimental data are shown in figure \ref{fig:distribution}
and table \ref{tab:average}. Except for $q_{0}$ and $V_{loop}$,
the average similarity of other key signals is greater than 90\%.
And the similarity distribution is concentrated above 90\%. The average
similarity of $q_{0}$ is greater than 85\%. This quantity has a poor
similarity, because if the equilibrium reconstruction is not properly
constrained with pressure profiles and kinetic measurements, $q_{0}$
might be unreliable. Since it might suffer from large variance, as
the model is struggling mostly for the reconstruction of that parameter.
Most of the errors in the reconstruction of $V_{loop}$ are related
to the plasma start-up and shut-down phases. To sum up, all selected
key diagnostic signals, excluding $V_{loop}$ are regarded as almost
wholly predicted.

\section{Conclusion and discussion \label{sec:Conclusion}}

In the present work, we propose a tokamak discharge prediction method
based on bidirectional LSTM. The bidirectional LSTM was developed
to introduce the wider contextual information of discharge sequence
to achieve a more efficient model. The model was trained on the EAST
experimental dataset in shot range \#14866-88283. This model can use
the actuator signals to reproduce the normal discharge evolution process
of the eleven key signals (i.e., actual plasma current $I_{p}$, normalized
beta $\beta_{n}$, toroidal beta $\beta_{t}$, beta poloidal $\beta_{p}$,
electron density $n_{e}$, store energy $W_{mhd}$, loop voltage $V_{loop}$,
elongation at plasma boundary $\kappa$, internal inductance $l_{i}$,
q at magnetic axis $q_{0}$, and q at 95\% flux surface $q_{95}$.)
without having as input any quantity derived by physical models. Bidirectional
LSTM architecture is robust to outliers. The average similarities
of all the selected key diagnostic signals between modeling results
and the experimental data are greater than 90\%, except for the $V_{loop}$
and $q_{0}$. The results presented in this paper demonstrate the
effectiveness of using a purely data-driven model to assist the validation
of the experimental proposal for a tokamak discharge.

The present work demonstrates that the model can easily be extended
to more diagnostic signals. Compared with physical models, experimental
data-driven models have proven to be very efficient computationally.
Once the machine learning model has been trained, a run of the trained
machine learning model is faster by orders of magnitude with respect
to the whole process of tokamak discharge modeling. The total inference
time of our model for an entire discharge prediction is about 1s.
The present work shows very promising results exploiting experimental
data-driven modeling as a supplement to physical-driven modeling tokamak.
Another important point which is worth mentioning is that our model
currently uses a smoothed version of the measured actuator signals
as input. and not directly the corresponding programmed PCS trajectories.
A complete understanding of the actuator trajectories prior to a shot
is challenging. The feasibility of using programmed PCS trajectories
as model input to predict the evolution of a tokamak discharge will
be the object of future work. Besides, we want to integrate the model
into the plasma control system (PCS) to automatically check the control
strategy. 1D and 2D plasma profiles (kinetic quantities, radiation
distribution) are also particularly important for tokamak discharge
modeling, since they can support scenario development with particular
reference to operational limits in high-performance scenarios \citep{Pau2019}.

\ack{}{}

The authors would like to thank all the members of EAST Team for providing
such a large quantity of past experimental data. The authors sincerely
thank Prof. Qiping Yuan, Dr. Ruirui Zhang, and Prof. Jinping Qian
for explaining the experimental data.

This work was supported by the National Key R\&D project under Contract
No.Y65GZ10593, the National MCF Energy R\&D Program under Contract
No.2018YFE0304100, and the Comprehensive Research Facility for Fusion
Technology Program of China under Contract No. 2018-000052-73-01-001228.

\bibliographystyle{unsrt}
\bibliography{library}

\begin{thebibliography}{10}

\bibitem{Falchetto2014}
G.L. Falchetto, David Coster, Rui Coelho, B.D. Scott, Lorenzo Figini, Denis
  Kalupin, Eric Nardon, Silvana Nowak, L.L. Alves, J.F. Artaud, V.~Basiuk,
  Jo{\~{a}}o~P.S. Bizarro, C.~Boulbe, A.~Dinklage, D.~Farina, B.~Faugeras,
  J.~Ferreira, A.~Figueiredo, Ph~Huynh, F.~Imbeaux, I.~Ivanova-Stanik,
  T.~Jonsson, H.-J. Klingshirn, C.~Konz, A.~Kus, N.B. Marushchenko,
  G.~Pereverzev, M.~Owsiak, E.~Poli, Y.~Peysson, R.~Reimer, J.~Signoret,
  O.~Sauter, R.~Stankiewicz, P.~Strand, I.~Voitsekhovitch, E.~Westerhof,
  T.~Zok, and W.~Zwingmann.
\newblock {The European Integrated Tokamak Modelling (ITM) effort: achievements
  and first physics results}.
\newblock {\em Nuclear Fusion}, 54(4):043018, apr 2014.

\bibitem{Bonoli2015}
Paul Bonoli, Lois~Curfman McInnes, C~Sovinec, D~Brennan, T~Rognlien, P~Snyder,
  J~Candy, C~Kessel, J~Hittinger, and L~Chacon.
\newblock {Report of the Workshop on Integrated Simulations for Magnetic Fusion
  Energy Sciences}.
\newblock Technical report, Massachusetts Institute of Technology, 2015.

\bibitem{kates-harbeck2019be}
Julian Kates-Harbeck, Alexey Svyatkovskiy, and William Tang.
\newblock {Predicting disruptive instabilities in controlled fusion plasmas
  through deep learning}.
\newblock {\em Nature}, 568(7753):526--531, apr 2019.

\bibitem{Hu2021}
W.H.~H. Hu, Cristina Rea, Q.P.~P. Yuan, K.G.~G. Erickson, D.L.~L. Chen, Biao
  Shen, Yao Huang, J.Y.~Y. Xiao, J.J.~J. Chen, Y.M.~M. Duan, Yang Zhang,
  H.D.~D. Zhuang, J.C.~C. Xu, K.J.~J. Montes, R.S.~S. Granetz, Long Zeng,
  J.P.~P. Qian, B.J.~J. Xiao, and J.G.~G. Li.
\newblock {Real-time prediction of high-density EAST disruptions using random
  forest}.
\newblock {\em Nuclear Fusion}, 61(6):066034, jun 2021.

\bibitem{Guo2021}
Bihao~H Guo, Dalong~L Chen, Biao Shen, Cristina Rea, Robert~S Granetz, Long
  Zeng, Wenhui~H Hu, Jinping~P Qian, Youwen~W Sun, and Bingjia~J Xiao.
\newblock {Disruption prediction on EAST tokamak using a deep learning
  algorithm}.
\newblock {\em Plasma Physics and Controlled Fusion}, 63(11):115007, nov 2021.

\bibitem{Cannas2007}
B.~Cannas, A.~Fanni, P.~Sonato, and M.K. Zedda.
\newblock {A prediction tool for real-time application in the disruption
  protection system at JET}.
\newblock {\em Nuclear Fusion}, 47(11):1559--1569, nov 2007.

\bibitem{cannas2007support}
Barbara Cannas, R.S. Delogu, ALESSANDRA Fanni, P~Sonato, and M.K. Zedda.
\newblock {Support vector machines for disruption prediction and novelty
  detection at JET}.
\newblock {\em Fusion Engineering and Design}, 82(5-14):1124--1130, oct 2007.

\bibitem{Cannas2010}
Barbara Cannas, Alessandra Fanni, G~Pautasso, G~Sias, and P~Sonato.
\newblock {An adaptive real-time disruption predictor for ASDEX Upgrade}.
\newblock {\em Nuclear Fusion}, 50(7):075004, jul 2010.

\bibitem{Yoshino2003}
R~Yoshino.
\newblock {Neural-net disruption predictor in JT-60U}.
\newblock {\em Nuclear Fusion}, 43(12):1771--1786, dec 2003.

\bibitem{clayton2013electron}
D.~J. Clayton, K.~Tritz, D.~Stutman, R.~E. Bell, A.~Diallo, B~P LeBlanc, and
  M.~Podest{\`{a}}.
\newblock {Electron temperature profile reconstructions from multi-energy SXR
  measurements using neural networks}.
\newblock {\em Plasma Physics and Controlled Fusion}, 55(9):095015, sep 2013.

\bibitem{Barana2002}
O~Barana, A~Murari, P~Franz, L~C Ingesson, and G~Manduchi.
\newblock {Neural networks for real time determination of radiated power in
  JET}.
\newblock {\em Review of Scientific Instruments}, 73(5):2038--2043, may 2002.

\bibitem{CANNAS2019374}
B~Cannas, S~Carcangiu, A~Fanni, T~Farley, F~Militello, A~Montisci, F~Pisano,
  G~Sias, and N~Walkden.
\newblock {Towards an automatic filament detector with a Faster R-CNN on
  MAST-U}.
\newblock {\em Fusion Engineering and Design}, 146:374--377, sep 2019.

\bibitem{Honda2019}
M.~Honda and E.~Narita.
\newblock {Machine-learning assisted steady-state profile predictions using
  global optimization techniques}.
\newblock {\em Physics of Plasmas}, 26(10):102307, oct 2019.

\bibitem{Meneghini2017}
O~Meneghini, S~P Smith, P~B Snyder, G~M Staebler, J~Candy, E~Belli, L~Lao,
  M~Kostuk, T~Luce, T~Luda, J~M Park, and F~Poli.
\newblock {Self-consistent core-pedestal transport simulations with neural
  network accelerated models}.
\newblock {\em Nuclear Fusion}, 57(8):86034, jul 2017.

\bibitem{Meneghini2021}
O.~Meneghini, G.~Snoep, B.C. Lyons, J.~McClenaghan, C.S. Imai, B.~Grierson,
  S.P. Smith, G.M. Staebler, P.B. Snyder, J.~Candy, E.~Belli, L.~Lao, J.M.
  Park, J.~Citrin, T.L. Cordemiglia, A.~Tema, and S.~Mordijck.
\newblock {Neural-network accelerated coupled core-pedestal simulations with
  self-consistent transport of impurities and compatible with ITER IMAS}.
\newblock {\em Nuclear Fusion}, 61(2):026006, feb 2021.

\bibitem{Murari2012}
A~Murari, D~Mazon, N~Martin, G~Vagliasindi, and M~Gelfusa.
\newblock {Exploratory Data Analysis Techniques to Determine the Dimensionality
  of Complex Nonlinear Phenomena: The L-to-H Transition at JET as a Case
  Study}.
\newblock {\em IEEE Transactions on Plasma Science}, 40(5):1386--1394, may
  2012.

\bibitem{Ferreira2020}
Diogo~R Ferreira and Pedro~J Carvalho.
\newblock {Deep Learning for Plasma Tomography in Nuclear Fusion}.
\newblock pages 1--5, 2020.

\bibitem{MURARI20132}
A~Murari, P~Arena, A~Buscarino, L~Fortuna, and M~Iachello.
\newblock {On the identification of instabilities with neural networks on JET}.
\newblock {\em Nuclear Instruments and Methods in Physics Research Section A:
  Accelerators, Spectrometers, Detectors and Associated Equipment}, 720:2--6,
  aug 2013.

\bibitem{Boyer2019}
M.D. Boyer, S.~Kaye, and K.~Erickson.
\newblock {Real-time capable modeling of neutral beam injection on NSTX-U using
  neural networks}.
\newblock {\em Nuclear Fusion}, 59(5):056008, may 2019.

\bibitem{MURARI2010850}
A~Murari, J~Vega, D~Mazon, D~Patan{\'{e}}, G~Vagliasindi, P~Arena, N~Martin,
  N~F Martin, G~Ratt{\'{a}}, and V~Caloone.
\newblock {Machine learning for the identification of scaling laws and
  dynamical systems directly from data in fusion}.
\newblock {\em Nuclear Instruments and Methods in Physics Research Section A:
  Accelerators, Spectrometers, Detectors and Associated Equipment},
  623(2):850--854, 2010.

\bibitem{Gaudio_2014}
P~Gaudio, A~Murari, M~Gelfusa, I~Lupelli, and J~Vega.
\newblock {An alternative approach to the determination of scaling law
  expressions for the L{\{}$\backslash$textendash{\}}H transition in Tokamaks
  utilizing classification tools instead of regression}.
\newblock {\em Plasma Physics and Controlled Fusion}, 56(11):114002, oct 2014.

\bibitem{Bockenhoff2018}
Daniel B{\"{o}}ckenhoff, Marko Blatzheim, Hauke H{\"{o}}lbe, Holger Niemann,
  Fabio Pisano, Roger Labahn, and Thomas~Sunn Pedersen.
\newblock {Reconstruction of magnetic configurations in W7-X using artificial
  neural networks}.
\newblock {\em Nuclear Fusion}, 58(5):56009, mar 2018.

\bibitem{coccorese1994identification}
E~Coccorese, C~Morabito, and Raffaele Martone.
\newblock {Identification of noncircular plasma equilibria using a neural
  network approach}.
\newblock {\em Nuclear Fusion}, 34(10):1349--1363, oct 1994.

\bibitem{Bishop1994}
Chris~M Bishop, Paul~S Haynes, Mike E~U Smith, Tom~N Todd, and David~L Trotman.
\newblock {Fast feedback control of a high temperature fusion plasma}.
\newblock {\em Neural Computing {\&} Applications}, 2(3):148--159, sep 1994.

\bibitem{Jeon2001}
Young-Mu Jeon, Yong-Su Na, Myung-Rak Kim, and Y~S Hwang.
\newblock {Newly developed double neural network concept for reliable fast
  plasma position control}.
\newblock {\em Review of Scientific Instruments}, 72(1):513--516, jan 2001.

\bibitem{Wang2016a}
S~Y Wang, Z~Y Chen, D~W Huang, R~H Tong, W~Yan, Y~N Wei, T~K Ma, M~Zhang, and
  G~Zhuang.
\newblock {Prediction of density limit disruptions on the J-TEXT tokamak}.
\newblock {\em Plasma Physics and Controlled Fusion}, 58(5):055014, may 2016.

\bibitem{Joung2020}
Semin Joung, Jaewook Kim, Sehyun Kwak, J.~G. Bak, S.~G. Lee, H.~S. Han, H.~S.
  Kim, Geunho Lee, Daeho Kwon, and Y.-C.~C. Ghim.
\newblock {Deep neural network Grad-Shafranov solver constrained with measured
  magnetic signals}.
\newblock {\em Nuclear Fusion}, 60(1):16034, dec 2020.

\bibitem{Milligen1995}
B~Ph. van Milligen, V~Tribaldos, and J~A Jim{\'{e}}nez.
\newblock {Neural Network Differential Equation and Plasma Equilibrium Solver}.
\newblock {\em Phys. Rev. Lett.}, 75(20):3594--3597, nov 1995.

\bibitem{Bishop1995}
Chris~M Bishop, Paul~S Haynes, Mike E~U Smith, Tom~N Todd, and David~L Trotman.
\newblock {Real-time control of a tokamak plasma using neural networks}.
\newblock {\em Neural Computation}, 7(1):206--217, 1995.

\bibitem{Yang_2020}
Bin Yang, Zhenxing Liu, Xianmin Song, and Xiangwen Li.
\newblock {Design of {\{}HL{\}}-2A plasma position predictive model based on
  deep learning}.
\newblock {\em Plasma Physics and Controlled Fusion}, 62(12):125022, nov 2020.

\bibitem{Wakatsuki2019}
T.~Wakatsuki, T.~Suzuki, N.~Hayashi, N.~Oyama, and S.~Ide.
\newblock {Safety factor profile control with reduced central solenoid flux
  consumption during plasma current ramp-up phase using a reinforcement
  learning technique}.
\newblock {\em Nuclear Fusion}, 59(6):066022, jun 2019.

\bibitem{Rasouli2013}
H.~Rasouli, C.~Rasouli, and A.~Koohi.
\newblock {Identification and control of plasma vertical position using neural
  network in Damavand tokamak}.
\newblock {\em Review of Scientific Instruments}, 84(2):023504, feb 2013.

\bibitem{yang2020modeling}
Bin Yang, Zhenxing Liu, Xianmin Song, Xiangwen Li, and Yan Li.
\newblock {Modeling of the HL-2A plasma vertical displacement control system
  based on deep learning and its controller design}.
\newblock {\em Plasma Physics and Controlled Fusion}, 62(7):75004, jul 2020.

\bibitem{Seo2021}
Jaemin Seo, Y.-S. Na, B.~Kim, C.Y. Lee, M.S. Park, S.J. Park, and Y.H. Lee.
\newblock {Feedforward beta control in the KSTAR tokamak by deep reinforcement
  learning}.
\newblock {\em Nuclear Fusion}, 61(10):106010, oct 2021.

\bibitem{Mathews2021}
A.~Mathews, M.~Francisquez, J.~W. Hughes, D.~R. Hatch, B.~Zhu, and B.~N.
  Rogers.
\newblock {Uncovering turbulent plasma dynamics via deep learning from partial
  observations}.
\newblock {\em Physical Review E}, 104(2):025205, aug 2021.

\bibitem{wan2021}
Chenguang Wan, Zhi Yu, Feng Wang, Xiaojuan Liu, and Jiangang Li.
\newblock {Experiment data-driven modeling of tokamak discharge in EAST}.
\newblock {\em Nuclear Fusion}, 61(6):066015, jun 2021.

\bibitem{Jardin_1993}
S~C Jardin, M~G Bell, N~Pomphrey, Home Search, Collections Journals, About
  Contact, My~Iopscience, I~P Address, S~C Jardin, M~G Bell, and N~Pomphrey.
\newblock {TSC simulation of Ohmic discharges in TFTR}.
\newblock {\em Nuclear Fusion}, 33(3):371--382, mar 1993.

\bibitem{Jardin1986}
S.C Jardin, N~Pomphrey, and J~Delucia.
\newblock {Dynamic modeling of transport and positional control of tokamaks}.
\newblock {\em Journal of Computational Physics}, 66(2):481--507, oct 1986.

\bibitem{Wan2015}
Baonian Wan, Jiangang Li, Houyang Guo, Yunfeng Liang, Guosheng Xu, Liang Wang,
  and Xianzu Gong.
\newblock {Advances in H-mode physics for long-pulse operation on
  {\{}EAST{\}}}.
\newblock {\em Nuclear Fusion}, 55(10):104015, jul 2015.

\bibitem{Wan2013}
Baonian Wan, Jiangang Li, Houyang Guo, Yunfeng Liang, and Guosheng Xu.
\newblock {Progress of long pulse and H-mode experiments in EAST}.
\newblock {\em Nuclear Fusion}, 53(10):104006, oct 2013.

\bibitem{Li2011}
Jiangang Li and Baonian Wan.
\newblock {Recent progress in RF heating and long-pulse experiments on EAST}.
\newblock {\em Nuclear Fusion}, 51(9):094007, sep 2011.

\bibitem{thireou2007}
Trias Thireou and Martin Reczko.
\newblock {Bidirectional Long Short-Term Memory Networks for Predicting the
  Subcellular Localization of Eukaryotic Proteins}.
\newblock {\em IEEE/ACM Transactions on Computational Biology and
  Bioinformatics}, 4(3):441--446, jul 2007.

\bibitem{Schuster1997}
Mike Schuster and K.K. Paliwal.
\newblock {Bidirectional recurrent neural networks}.
\newblock {\em IEEE Transactions on Signal Processing}, 45(11):2673--2681,
  1997.

\bibitem{Graves2005a}
Alex Graves and J{\"{u}}rgen Schmidhuber.
\newblock {Framewise phoneme classification with bidirectional LSTM and other
  neural network architectures}.
\newblock {\em Neural Networks}, 18(5-6):602--610, jul 2005.

\bibitem{Abduljabbar2021}
Rusul~L. Abduljabbar, Hussein Dia, and Pei-Wei Tsai.
\newblock {Development and evaluation of bidirectional LSTM freeway traffic
  forecasting models using simulation data}.
\newblock {\em Scientific Reports}, 11(1):23899, dec 2021.

\bibitem{Siami-Namini2019}
Sima Siami-Namini, Neda Tavakoli, and Akbar~Siami Namin.
\newblock {The Performance of LSTM and BiLSTM in Forecasting Time Series}.
\newblock In {\em 2019 IEEE International Conference on Big Data (Big Data)},
  pages 3285--3292. IEEE, dec 2019.

\bibitem{Srivastava2014}
Nitish Srivastava, Geoffrey Hinton, Alex Krizhevsky, Ilya Sutskever, and Ruslan
  Salakhutdinov.
\newblock {Dropout: A simple way to prevent neural networks from overfitting}.
\newblock {\em Journal of Machine Learning Research}, 15(1):1929--1958, 2014.

\bibitem{wang2018studyof}
Feng Wang, Yueting Wang, Ying Chen, Shi Li, and Fei Yang.
\newblock {Study of web-based management for EAST MDSplus data system}.
\newblock {\em Fusion Engineering and Design}, 129(June 2017):88--93, apr 2018.

\bibitem{DeTommasi2019}
Gianmaria {De Tommasi}.
\newblock {Plasma Magnetic Control in Tokamak Devices}.
\newblock {\em Journal of Fusion Energy}, 38(3-4):406--436, aug 2019.

\bibitem{Anand2021b}
H~Anand, D~Humphreys, D~Eldon, A~Leonard, A~Hyatt, B~Sammuli, and A~Welander.
\newblock {Plasma flux expansion control on the DIII-D tokamak}.
\newblock {\em Plasma Physics and Controlled Fusion}, 63(1):015006, jan 2021.

\bibitem{mongodb}
{MongoDB main page}.

\bibitem{Dean2008}
Jeffrey Dean and Sanjay Ghemawat.
\newblock {MapReduce}.
\newblock {\em Communications of the ACM}, 51(1):107--113, jan 2008.

\bibitem{dean2012large}
Jeffrey Dean, Greg~S Corrado, Rajat Monga, Kai Chen, Matthieu Devin, Quoc~V Le,
  Mark~Z Mao, Marc'Aurelio~Marc'aurelio Ranzato, Andrew Senior, Paul Tucker,
  Others, and Ke~Yang.
\newblock {Large scale distributed deep networks}.
\newblock {\em Advances in neural information processing systems},
  25:1223--1231, 2012.

\bibitem{Huang2013}
Zhiheng Huang, Geoffrey Zweig, Michael Levit, Benoit Dumoulin, Barlas Oguz, and
  Shawn Chang.
\newblock {Accelerating recurrent neural network training via two stage classes
  and parallelization}.
\newblock In {\em 2013 IEEE Workshop on Automatic Speech Recognition and
  Understanding}, pages 326--331. IEEE, dec 2013.

\bibitem{Chetlur2014}
Sharan Chetlur, Cliff Woolley, Philippe Vandermersch, Jonathan Cohen, John
  Tran, Bryan Catanzaro, and Evan Shelhamer.
\newblock {cuDNN: Efficient Primitives for Deep Learning}.
\newblock oct 2014.

\bibitem{Khomenko2016}
Viacheslav Khomenko, Oleg Shyshkov, Olga Radyvonenko, and Kostiantyn Bokhan.
\newblock {Accelerating recurrent neural network training using sequence
  bucketing and multi-GPU data parallelization}.
\newblock In {\em 2016 IEEE First International Conference on Data Stream
  Mining {\&} Processing (DSMP)}, pages 100--103. IEEE, aug 2016.

\bibitem{Xiao2008}
B.J. Xiao, D.A. Humphreys, M.L. Walker, A.~Hyatt, J.A. Leuer, D.~Mueller, B.G.
  Penaflor, D.A. Pigrowski, R.D. Johnson, A.~Welander, Q.P. Yuan, H.Z. Wang,
  J.R. Luo, Z.P. Luo, C.Y. Liu, L.Z. Liu, and K.~Zhang.
\newblock {EAST plasma control system}.
\newblock {\em Fusion Engineering and Design}, 83(2-3):181--187, apr 2008.

\bibitem{NEURIPS2019_9015}
Adam Paszke, Sam Gross, Francisco Massa, Adam Lerer, James Bradbury, Gregory
  Chanan, Trevor Killeen, Zeming Lin, Natalia Gimelshein, Luca Antiga, Alban
  Desmaison, Andreas Kopf, Edward Yang, Zachary DeVito, Martin Raison, Alykhan
  Tejani, Sasank Chilamkurthy, Benoit Steiner, Lu~Fang, Junjie Bai, and Soumith
  Chintala.
\newblock {PyTorch: An Imperative Style, High-Performance Deep Learning
  Library}.
\newblock In H~Wallach, H~Larochelle, A~Beygelzimer,
  F~d$\backslash$textquotesingle Alch{\'{e}}-Buc, E~Fox, and R~Garnett,
  editors, {\em Advances in Neural Information Processing Systems 32}, pages
  8024--8035. Curran Associates, Inc., 2019.

\bibitem{Glorot2010}
Xavier Glorot and Yoshua Bengio.
\newblock {Understanding the difficulty of training deep feedforward neural
  networks}.
\newblock In {\em Journal of Machine Learning Research}, volume~9, pages
  249--256, 2010.

\bibitem{Graves2013}
Alex Graves.
\newblock {Generating Sequences With Recurrent Neural Networks}.
\newblock aug 2013.

\bibitem{Pau2019}
A.~Pau, A.~Fanni, S.~Carcangiu, B.~Cannas, G.~Sias, A.~Murari, F.~Rimini,
  Human~Immunodeficiency Virus, Associated~Neurocognitive Disorders, Consensus
  Report, Mind~Corresponding Author, and Alternate~Corresponding Author.
\newblock {A machine learning approach based on generative topographic mapping
  for disruption prevention and avoidance at JET}.
\newblock {\em Nuclear Fusion}, 59(10):106017, oct 2019.

\end{thebibliography}

\end{document}